\documentclass[11pt]{article}

\usepackage{epsfig}
\usepackage{amssymb}
\usepackage{amsthm}
\usepackage{amsmath}
\usepackage{amssymb,amsfonts}
\usepackage{graphicx}
\newcommand{\be}{\begin{eqnarray}}
\newcommand{\ee}{\end{eqnarray}}
\renewcommand{\d}{\partial}

\def\lsim{\mathrel{\rlap {\raise.5ex\hbox{$ < $}}
{\lower.5ex\hbox{$\sim$}}}}
\def\gsim{\mathrel{\rlap {\raise.5ex\hbox{$ > $}}
{\lower.5ex\hbox{$\sim$}}}} \topmargin -1.5cm \textheight=22.5cm
\begin{document}

\begin{titlepage}

\begin{centering}

\hfill hep-th/0403029\\

\vspace{1 in}
{\bf {BOX COMPACTIFICATION AND SUPERSYMMETRY BREAKING } }\\
\vspace{2 cm} {A. KEHAGIAS$^{1}$
 and K. TAMVAKIS$^{2}$}\\
\vskip 1cm
{$^1 $\it{Physics Department, National Technical University\\
15 780 Zografou, Athens,  GREECE}\\
\vskip 1cm
{$^2$\it {Physics Department, University of Ioannina\\
45110 Ioannina, GREECE}}}\\

\vspace{1.5cm}
{\bf Abstract}\\
\end{centering}
\vspace{.1in}
We discuss all possible compactifications on flat three-dimensional  spaces.
In particular, various fields are studied on a  box with opposite sides
identified, after two of them are rotated by $\pi$, and their spectra are obtained.
The compactification of a general $7D$ supersymmetric theory in such a box is considered and the
corresponding four-dimensional theory is studied, in relation to the boundary
conditions chosen. The resulting  spectrum,
according to the allowed field boundary conditions, corresponds to partially or completely  broken supersymmmetry.
We briefly discuss also  the
breaking of gauge symmetries under the proposed box compactification.

\vfill

\vspace{2cm}
\begin{flushleft}

March 2004
\end{flushleft}
\hrule width 6.7cm \vskip.1mm{\small \small}
 \end{titlepage}

In almost all extensions beyond the Standard Model, supersymmetry plays a central role. In particular,
Superstring Theory\cite{Green}, as well as related theories of extended objects \cite{Polchinski},
provide a framework for a quantum theory of gravity. Nevertheless, since supersymmetry is not
a low-energy symmetry of Nature, and has to be broken, supersymmetry
breaking should be a key ingredient of the final theory. This important issue is still open.
The tree-level  {\textit{Scherk-Schwarz Supersymmetry Breaking}} (SSSB)
 mechanism \cite{SSSB}--\cite{Antoniadis:1998sd} is one of the proposals put forward, linking supersymmetry breaking to compactification. The smallness of supersymmetry breaking scale in comparison to the other
 scales, like the traditional unification or Planck scales, if it is to be associated with compactification, requires the presence of
 large extra dimensions\cite{Antoniadis:1990ew},\cite{Antoniadis:1997zg}. Many models
 of this type have been proposed in the last few years \cite{Antoniadis:1998sd} and, although, none is
 phenomenologically waterproof, it is generally admitted that the possibility of extra dimensions
 at the $TeV$ scale is open. In SSSB one takes advantage of the R-symmetry of the supersymmetric theory
 to shift appropriatelly the masses of  bosons and fermions lifting in this way the degeneracy and, thus, breaking supersymmetry.
 Alternative ways of breaking supersymmetry include gaugino condensation in the hidden sector \cite{Amati} or, in brane scenarios \cite{Dudas},
 bulk to brane  and brane to
brane supersymmetry breaking \cite{Rattazzi}.   Supersymmetry
 may also be broken by background fluxes \cite{Bachas},\cite{Vafa}. In the case of background magnetic fields,
  the occuring tadpoles of which, will presummably be removed in the full quantum theory
 \cite{Bachas}.

 In the present short article we elaborate on the possibility of breaking supersymmetry at the compactification process employing a novel
 compactification scheme. Gauge symmetry breaking as a result of compactification is also studied. Thus, as far as supersymmetry breaking is
 concerned, although we work along
 the lines of SSSB, it should be stressed that there is a fundamental difference with it, since in SSSB the boundary
  conditions for R-symmetry singlets, like vectors, are always periodic,
  in contrast to our {\textit{box compactification}}, where they can be non-trivial even for R-singlets.
   In addition to that, the profile of
  our supersymmetry breaking is always that of a vanishing supertrace,
  resembling spontaneous breaking, in contrast to the SSSB patterns.
   We shall discuss our main differences with SSSB later on.
At the moment, let us recall that according to a theoretical proposal,
we are living in  a $4+n$-dimensional space-time, $n$ dimensions of which have been
compactified to form a
orientable  compact space $X^n$. By turning off all fields except gravity, Einstein equations require the vacuum to be Ricci-flat and,
thus, it is of the form $M^4\times X^n$, where $M^4$ is the four dimensional Minkowski space-time.
The internal manifold $X^n$ is assumed to be a complete, connected and
compact  Ricci-flat manifold like a Calabi-Yau space (in the case of String Theory). Nevertheless, one may
assume that{\textit{ $X^n$ is flat and not just Ricci-flat}}. In that case, the possible vacua are
orientable compact euclidean space-forms. The most well studied case is that of an $n$-dimensional torus $T^n$. Other cases
involve orbifolds of $T^n$ by some discrete group, which although are singular spaces, strings can consistently propagate on them.
These kind of orbifolds can also be obtained as limiting cases of smooth Calabi-Yau space. In this case, all curvature of the Calabi-Yau space
is concentrated at the orbifold points. However, here we shall be interested in smooth, compact and flat $n$-dimensional spaces

 Unfortunatelly, existing classifications \cite{Wolf} of orientable compact euclidean space-forms do not go beyond $3D$. In particular, in two
 dimensions, the only orientable
compact euclidean space-form is the torus $T^2$. In three
dimensions we have the following possibilities by making
identifications on possible fundamental polyhedra in
$\mathbb{R}^3$:

\begin{figure}[!h]
\centerline{\hbox{\psfig{figure=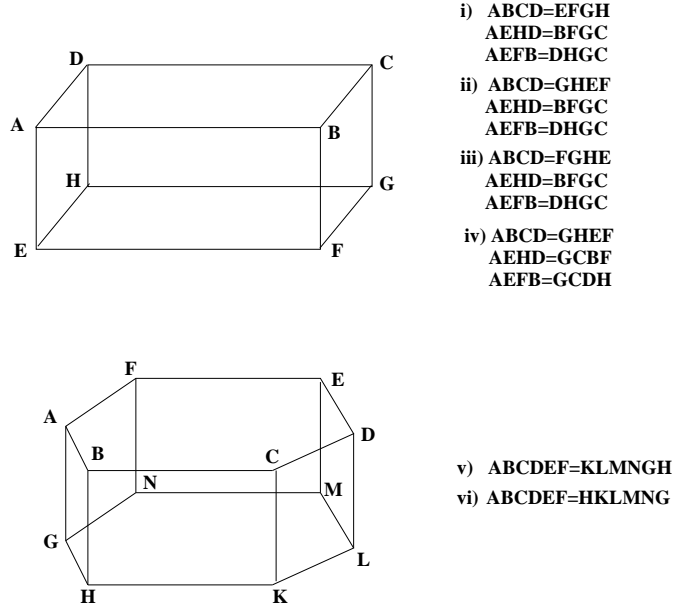,height=8cm}}}
\caption{ {\footnotesize \em Possible identification on $\mathbb{R}^3$ which produce compact orientable three-spaces.
}}
\label{f1a}
\end{figure}


{\allowdisplaybreaks
\begin{description}

\item[i)] On a paralepiped by identifying opposite sides,
\vspace{-.3cm}
\item[ii)] On a paralepiped by identifying opposite sides, one pair rotated by $\pi$,
\vspace{-.3cm}
\item[iii)] On a paralepiped by identifying opposite sides, one pair rotated by $\pi/2$,
\vspace{-.3cm}
\item[iv)] On a paralepiped by identifying opposite sides, all pairs rotated by $\pi$.
\vspace{-.3cm}
\item[v)] On a hexagonal prism by identifying opposite sides, the top  rotated by $2\pi/3$ with respect to the bottom,
\vspace{-.3cm}
\item[vi)] On a hexagonal prism by identifying opposite sides, the top  rotated by $\pi/3$ with respect to the bottom.
\end{description}}

In addition to the above, there exist four non-compact orientable  euclidean space-forms, four non-compact and
non-orientable and four compact and non-orientable euclidean space-forms. This makes a total of 18 distinct types of
 locally euclidean spaces. Of them, only $\mathbb{R}^3$  is simply connected while the rest of the spaces are connected
 to the 17 crystallographic groups. It should be noted that the non-orientable cases are obtained by including ``glide reflections",
 i.e. a reflection in a plane through the origin followed by a translation parallel to the plane.

In what follows we will assume a $7D$ theory which is spontaneously compactified to $4D$ on a  compact and smooth internal space.
According to the above discussion then, any flat $7D$ vacuum will be of the form $M^4\times X^3$, where $X^3$ is
any of the spaces $\rm{(i)-(vi)}$. One may easily recognize
that $\rm{(i)}$ is just $T^3$ while the rest of the cases are orbifolds of $T^3$ by a freely acting isometry.


To make the discussion concrete let us assume that the internal space is the
$3D$ box which is obtained after having identified its opposite sides with one
pair rotated by $\pi$, i.e, the case $(ii)$ on $\mathbb{R}^3$ with coordinates $(x,y,z)$ subject to the identifications
\be
&&(x,y,z)\thickapprox(x+ R_1,y,z)\nonumber \\
&&(x,y,z)\thickapprox(x,y+R_2,z)\nonumber\\
&&(x,y,z)\thickapprox(-x,-y,z+R_3)\, .\label{1}
\ee
So, we have the normal identifications under translations in the $x,y$ directions,
while points in the $z$ directions
are identified after a $\pi$-rotation in the perpendicular $x,\,y$ plane. We will call this space $B^3$. Corresponding efforts
for compactifications on squares \cite{Dobrescu} produce orbifold singularities.
There is a $\mathbb{Z}_2$ symmetry, which acts as on the coordinates as\footnote{The space $B^3$ may be viewed as
$T^3/\mathbb{Z}_2$. It is not an orbifold as
$\mathbb{Z}_2$ acts freely on $T^3$ (there are no fixed points under the action of $\mathbb{Z}_2$).}
\be
g: (x^1,x^2,x^3)\thickapprox(-x^1,-x^2,x^3+ R_3)
\ee
We observe that $g^2=1$ since
\be
g^2: (x^1,x^2,x^3)\thickapprox(x^1,x^2,x^3+2R_3)
\ee
and $(x,y,z)$, $(x,y,z+ 2R_3)$ are identified. Thus, $B^3$ is a double cover of $T^3$.

After having defined the geometry, we are now ready to study the behaviour
of fields in the  box of eq.(\ref{1}). It should be noted that we are mainly interested in the
$k_3$-periodicity as the periodicity  in $k_1,k_2$ are determined as usual by the identification
$x\thickapprox x+R_1,~~ y\thickapprox y+R_2$.
\begin{enumerate}
\item{ {\bf Scalar}}\\
A scalar field $\Phi$ is periodic on $T^3$ and on $B^3$. It should, therefore, satisfy
\be
\Phi(x^1,x^2,x^3)=\alpha \, \Phi(-x^1,-x^2,x^3+ R_3)=
\alpha^2\, \Phi(x^1,x^2,x^3+2R_3)
\ee
so that $\alpha^2=1$. Thus, on $B^3$, a scalar field may
have periodic or antiperiodic boundary conditions, i.e.,
\be
\Phi(x^1,x^2,x^3)=\pm \Phi(-x^1,-x^2,x^3+ R_3) \label{bd}
\ee
The eigenvalues of the scalar Laplace  operator $\nabla^2=
-\partial_i\partial^i$
 on $B^3$ are as usual $k^2=k_1^2+k_2^2+k_3^2$ and the corresponding
eigenstates $\cos(k_1x^1)\cos(k_2x^2)e^{i k_3x^3}$.
As $x^1,x^2$ are periodic with periods
 $R_1,~R_2$, respectively, we will always have (for the first eigenstates)
 \be
 k_1=\frac{2\pi n_1}{R_1}, ~~~~~k_2=\frac{2\pi n_2}{R_2}\, ,
~~~~~n_1,n_2=0,1,\ldots
 \ee
 On the other hand, the value of $k_3$ depends on the boundary
conditions (\ref{bd}). In particular we get
 \be
 &&k_{3}^{(+)}=  \frac{2 \pi n_3}{R_3}\, ,~~~~~ k_{3}^{(-)}=
\frac{(2n_3\!+\!1)\pi}{R_3}\, , ~~~n_3=0,1,\ldots
\ee
for the periodic $(+)$ and anti-periodic $(-)$ choice, respectively.
\item{{\bf Fermion}}

Similarly, for a fermion  $\Psi$ we should have
\be
\Psi(x^1,x^2,x^3)=\beta \, e^{ i\,  \phi \, \sigma_3}\,
\Psi(-x^1,-x^2,x^3+R_3)=
\beta^2\,e^{2i \phi \, \sigma_3} \Psi(x^1,x^2,x^3,x^3+2R_3) \label{Psi}
\ee
where $\sigma_3$ is a Pauli matrix.
For periodic $\Psi$ on $T^3$ we get that $\beta^2\,e^{2 i\, \phi \,
\sigma_3}=1$ so that
$\beta=\pm 1, ~ \phi=\pi$. Therefore, the boundary conditions for fermion fields
on $B^3$  are
\be
\Psi(x^1,x^2,x^3)=\pm \, e^{i \pi  \, \sigma_3}\, \Psi(-x^1,-x^2,x^3+ R_3)
\ee
and we get
\be
k_{3}^{(+)}=  \frac{2 \pi n_3}{R_3}+\frac{\pi}{R_3}\sigma_3
\, , ~~~~~~ k_{3}^{(-)}=
\frac{2 \pi n_3}{R_3}+\frac{\pi}{R_3}(1+\sigma_3)
\ee
Clearly, the ``periodic" $(+)$  condition makes the fermion massive with mass $m^2=\pi^2/R_3^2$.
In contrast, the second, ``anti-periodic" $(-)$, boundary condition, due to the projection
operator $(1+\sigma_3)$, makes the upper component of $\Psi$
massive, while its lower component has a zero mode.

\item{{\bf Vector}}

For a vector $A_i$ we will have
\be
A_i(x^1,x^2,x^3)&=&\gamma\, \left(e^{i\, \theta \, J_3}\right)_i^j
A_j(x^1,x^2,x^3+ R_3)\nonumber \\
&=&
\gamma^2\, \left(e^{i\,  \theta \, J_3}\right)_i^j
\left(e^{ i\, \theta \, J_3}\right)_j^k A_k(x^1,x^2,x^3+2  R_3)
\ee
where $J_3={\rm diag}(\sigma_2,0)$ is the generator of rotations
in the $x^1,\,x^2$ plane and so
\be
\gamma^2\, \left(e^{i\,  \theta \, J_3}\right)_i^j
\left(e^{ i\, \theta \, J_3}\right)_j^k=\delta_i^k
\ee
It is not difficult then to verify that $\theta=\pi$ and
\be
A_i(x^1,x^2,x^3)&=&\pm \, R_i^j A_j(-x^1,-x^2,x^3+R_3)
\ee
where $R={\rm diag}(-1,-\sigma_3)$. Then, the eigenvalues for the
components of $A_i$ should be
\be
A_1,A_2&:&k_{3}^{(+)}=   \frac{(2n_3\!+\!1)\pi}{R_3}\, , ~~~~k_{3}^{(-)}=
  \frac{2 \pi n_3}{R_3}\\
A_3&:&k_{3}^{(+)}= \frac{2 \pi n_3}{R_3}\, , ~~~~k_{3}^{(-)}=
 \frac{(2n_3\!+\!1)\pi}{R_3}\\
\ee
for the periodic $(+)$ and antiperiodic $(-)$ boundary conditions, respectively.

\item{{\bf Symmetric two-tensor}}

For a symmetric two-tensor $h_{ij}$ we will have
\be
h_{ij}(x^1,x^2,x^3)&=&\pm\, R_i^\ell\,  R_j^k\, h_{ij}(-x^1,-x^2,x^3+R_3) \label{hh}
\ee
As a result, its $k_3$  eigenvalues will be
\be
h_{ij} \, (i,j\neq 3)\, , h_{33} &:&k_{3}^{(+)}=   \frac{(2n_3\!+\!1)\pi}{R_3}\, , ~~~~k_{3}^{(-)}=
  \frac{2 \pi n_3}{R_3}\\
h_{i3}\,(i\neq 3) &:&k_{3}^{(+)}= \frac{2 \pi n_3}{R_3}\, , ~~~~k_{3}^{(-)}=
 \frac{(2n_3\!+\!1)\pi}{R_3}\\
\ee
for the periodic $(+)$ and antiperiodic $(-)$ boundary conditions of eq.(\ref{hh}), respectively.
\end{enumerate}

It is clear that the components $A_1,A_2$ and $A_3$ of a vector $A_M$, as well as the components of a tensor, have different
$k_3$. This is due to the fact that the box we are employing here is a non-homogeneous space.


Let us now see how we can use the above to break Supersymmetry. We will consider
a $7D$ supersymmetric ${\cal{N}}=1$ theory \cite{Bergshoeff},\cite{Gherghetta} with a vector supermultiplet
which contains a vector $A_M$, 3 scalars $\phi^i\, , i=1,2,3$ and one
symplectic-Majorana spinor $\lambda^a\, , a=1,2$.
We would like to see the theory when we dimensionally reduce on the
space $B^3$. The effective $4D$ theory
then contains the following fields $(A_\mu, A_i, \phi^i,\lambda_1^a,\lambda_2^a)$,
i.e., a vector $A_\mu$, 6 scalars
$\Phi^I=(A_i,\phi^i), I=1,...6$ and 4 spinors
$\Psi^A=(\lambda_1^a,\lambda_2^a),A=1,...,4$. This is simply a vector
multiplet of a 4D ${\cal{N}}=4$ theory. All these fields depend
on the internal $x^1,x^2,x^3$ coordinates so we need to
expand in terms of harmonics on $B^3$.
 The harmonics
for the latter  are
\be
Y_{\{n_1n_2n_3\}}=\frac{1}{\sqrt{V}}\cos(k_1x^1)\cos(k_2 x^2) e^{i k_ix^i} \label{bs}
\ee
where $k_i=2\pi n_i/R_i, n_i=0,1,...$ and $V$ the volume of $B^3$.
Then, the expansion of the $4D$ fields is
\be
A_\mu=A_\mu(x)Y_{\{n\}}, ~~~ A_i=A_i(x)Y_{\{n\}}, ~~~
\phi=\phi(x)Y_{\{n\}}, ~~~\lambda^a=\lambda^a(x)Y_{\{n\}}
\ee
We have, thus, a tower of massive states with
the masses of the vectors, scalars and fermions given by
\be
&&M^2_V=k_1^2+k_2^2+k_3^2=\left(\frac{2\pi n_1}{R_1}\right)^2+
\left(\frac{2\pi n_2}{R_2}\right)^2+\left(\frac{2\pi n_3}{R_3}\right)^2\\
&&M_S^2=M_F^2=M_V^2
\ee
It can easily be checked that $Str M^2=0$.

For the box $(ii)$ we are considering, depending on the boundary
conditions, we have a basis
$$Y_{\{n\}}^{(\pm)}\Longrightarrow k_3^{(\pm)}$$
 as in
(\ref{bs}), but with $k_3=k_3^{(\pm)}$, respectively.
For instance, we may take for the bosons
\be
A_\mu=A_\mu(x)Y_{\{n\}}^{(+)}, ~~~ A_{1,2}=A_{1,2}(x)Y_{\{n\}}^{(-)},~~~A_{3}=A_{3}(x)Y_{\{n\}}^{(-)}
~~~ \phi^i=\phi^i(x)Y_{\{n\}}^{(-)}.
\ee
The corresponding mass spectrum is then
\be
\begin{array}{|c|c|c|} \hline
A_\mu & M^2_V=k_1^2+k_2^2+k_3^{(+)2}&\left(\frac{2\pi
n_1}{R_1}\right)^2+ \left(\frac{2\pi
n_2}{R_2}\right)^2+\left(\frac{2\pi n_3}{R_3}\right)^2 \\ \hline
A_{1,2}&
M^2_S=k_1^2+k_2^2+k_3^{(-)2}&\left(\frac{2\pi
n_1}{R_1}\right)^2+ \left(\frac{2\pi
n_2}{R_2}\right)^2+\left(\frac{2\pi n_3}{R_3}\right)^2\\\hline
A_3&M^2_S=k_1^2+k_2^2+k_3^{(-)2}&\left(\frac{2\pi
n_1}{R_1}\right)^2+ \left(\frac{2\pi
n_2}{R_2}\right)^2+\left(\frac{(2n_3\!+1\!)\pi}{R_3}\right)^2\\\hline
\phi^i&M^2_S=k_1^2+k_2^2+k_3^{(-)2}&\left(\frac{2\pi
n_1}{R_1}\right)^2+ \left(\frac{2\pi
n_2}{R_2}\right)^2+\left(\frac{(2n_3\!+1\!)\pi}{R_3}\right)^2\\\hline
\end{array} \label{A}
\ee

For the 7D spinors we recall that in $SO(7)\supset SU_L(2)\!\times\!SU_R(2)\!\times\!SU(2)$, we have
${\bf 8}=({\bf 2,1;2})+({\bf 1,2;2})$. As a result, a 7D spinor $\lambda$ is decomposed into two left and two right-handed
4D spinors. We may take
\be
~~~\lambda=\chi_L^\alpha(x)\otimes \epsilon^\alpha Y_{\{n\}}^{(-)}+\chi_R^\alpha(x)\otimes \theta^\alpha Y_{\{n\}}^{(-)}\, , ~~~\alpha=1,2
\ee
where $\epsilon^a\, ,\theta^a$ are  two-component spinors
and $\chi^a_{1,2}$ are  4D spinors. The mass spectrum of the 4D spinor is then
\be
\begin{array}{|c|c|c|} \hline
\chi^1_L& M^2_F=k_1^2+k_2^2+k_3^{(-)2}&\left(\frac{2\pi
n_1}{R_1}\right)^2+ \left(\frac{2\pi
n_2}{R_2}\right)^2+\left(\frac{n_3\pi}{R_3}\right)^2\\\hline
\chi^1_R& M^2_F=k_1^2+k_2^2+k_3^{(-)2}&\left(\frac{2\pi
n_1}{R_1}\right)^2+ \left(\frac{2\pi
n_2}{R_2}\right)^2+\left(\frac{2\pi n_3}{R_3}\right)^2\\ \hline

\chi^2_L& M^2_F=k_1^2+k_2^2+k_3^{(-)2}&\left(\frac{2\pi
n_1}{R_1}\right)^2+ \left(\frac{2\pi
n_2}{R_2}\right)^2+\left(\frac{(2n_3\!+1\!)\pi}{R_3}\right)^2\\\hline
\chi^2_R& M^2_F=k_1^2+k_2^2+k_3^{(-)2}&\left(\frac{2\pi
n_1}{R_1}\right)^2+ \left(\frac{2\pi
n_2}{R_2}\right)^2+\left(\frac{(2n_3\!+1\!)\pi}{R_3}\right)^2\\ \hline
\end{array} \label{x}
\ee
Thus, from tables (\ref{A},\ref{x}) we see that we get {\textit{one massless vector, two massless
scalars and two massless fermions of opposite chirality}},  all corresponding to $n_i=0$.
On the other hand, {\textit{four scalars and two spinors of opposite chirality do not have zero modes}}. The massless
spectrum in 4D is then a vector of a ${\cal{N}}=2$ theory.
As a result, compactification on  this particular box with the
above boundary conditions leads to the supersymmetry breaking
$${\cal{N}}=4\Longrightarrow {\cal {N}}=2$$ Note that the profile of the breaking is that of spontaneous supersymmetry breaking, since
the supertrace still vanishes.

A complete supersymmetry breaking can be also achieved by assuming the following expansion of the
7D spinor
\be
~~~\lambda=\chi_L^\alpha(x)\otimes \epsilon^\alpha Y_{\{n\}}^{(-)}+
\chi_R^\alpha(x)\otimes \theta^\alpha Y_{\{n\}}^{(+)}\, , ~~~\alpha=1,2 \label{ss}
\ee
In this case the spectrum of the $4D$ spinors  is
\be
\begin{array}{|c|c|c|} \hline
\chi^1_L& M^2_F=k_1^2+k_2^2+k_3^{(-)2}&\left(\frac{2\pi
n_1}{R_1}\right)^2+ \left(\frac{2\pi
n_2}{R_2}\right)^2+\left(\frac{n_3\pi}{R_3}\right)^2\\\hline
\chi^1_R& M^2_F=k_1^2+k_2^2+k_3^{(-)2}&\left(\frac{2\pi
n_1}{R_1}\right)^2+ \left(\frac{2\pi
n_2}{R_2}\right)^2+\left(\frac{(2n_3\!+1\!)\pi}{R_3}\right)^2\\ \hline

\chi^2_L& M^2_F=k_1^2+k_2^2+k_3^{(-)2}&\left(\frac{2\pi
n_1}{R_1}\right)^2+ \left(\frac{2\pi
n_2}{R_2}\right)^2+\left(\frac{(2n_3\!+1\!)\pi}{R_3}\right)^2\\\hline
\chi^2_R& M^2_F=k_1^2+k_2^2+k_3^{(-)2}&\left(\frac{2\pi
n_1}{R_1}\right)^2+ \left(\frac{2\pi
n_2}{R_2}\right)^2+\left(\frac{(2n_3\!+1\!)\pi}{R_3}\right)^2\\ \hline
\end{array} \label{xx}
\ee
We see that from tables (\ref{A},\ref{xx}) that the massless spectrum is a vector $A_\mu$, two scalars
$A_{1,2}$ and a left-handed 4D spinor, which is not-supersymmetric. Thus, adopting the expansion in eq.(\ref{ss}), we have completely break
supersymmetry $${\cal{N}}=4\Longrightarrow {\cal {N}}=0$$

We can also break ${\cal{N}}=4$ to ${\cal{N}}=1$ by considering different boundary conditions for the bosons of the 7D multiplet as well.
For example, let us take
\be
A_\mu=A_\mu(x)Y_{\{n\}}^{(+)}, ~~~ A_{1,2}=A_{1,2}(x)Y_{\{n\}}^{(-)},
A_{3}=A_{3}(x)Y_{\{n\}}^{(-)},
~~~ \phi^i=\phi^i(x)Y_{\{n\}}^{(-)}. \label{aa}
\ee
Then, the mass spectrum for the $4D$ fields is
\be
\begin{array}{|c|c|c|} \hline
A_\mu & M^2_V=k_1^2+k_2^2+k_3^{(-)2}&\left(\frac{2\pi
n_1}{R_1}\right)^2+ \left(\frac{2\pi
n_2}{R_2}\right)^2+\left(\frac{(2n_3\!+1\!)\pi}{R_3}\right)^2 \\ \hline
A_{1,2}&
M^2_S=k_1^2+k_2^2+k_3^{(-)2}&\left(\frac{2\pi
n_1}{R_1}\right)^2+ \left(\frac{2\pi
n_2}{R_2}\right)^2+\left(\frac{2\pi n_3}{R_3}\right)^2\\\hline
A_3&M^2_S=k_1^2+k_2^2+k_3^{(-)2}&\left(\frac{2\pi
n_1}{R_1}\right)^2+ \left(\frac{2\pi
n_2}{R_2}\right)^2+\left(\frac{(2n_3\!+1\!)\pi}{R_3}\right)^2\\\hline
\phi^i&M^2_S=k_1^2+k_2^2+k_3^{(-)2}&\left(\frac{2\pi
n_1}{R_1}\right)^2+ \left(\frac{2\pi
n_2}{R_2}\right)^2+\left(\frac{(2n_3\!+1\!)\pi}{R_3}\right)^2\\\hline
\end{array} \label{AA}
\ee
The massless sector then for the $4D$ fields expanded as in eqs.(\ref{ss},\ref{aa}) is given in tables (\ref{xx},\ref{AA}) and consists of two
scalars $A_{1,2}$ and one left-handed spinor. This is the massless representation of a chiral ${\cal{N}}=1$ supersymmetry.

We may also study the effective 4D theory after the DR over
$B_2=T^3/\mathbb{Z}_2$. Consider a 7D supersymmetric theory which
contains a a vector $A_M$, 3 scalars $\phi^i\, , i=1,2,3$ and one
symplectic-Majorana spinor $\lambda^a\, , a=1,2$, all in the
adjoint representation of a semisimple group G.
 After DR on $T^3$
with normal boundary conditions to 4D, the effective action turns
out to be \be S_{\rm eff}= &&\int d^4x {\rm
Tr}\left(-\frac{1}{4}F_{\mu\nu}F^{\mu\nu}-\frac{1}{2}\bar{\lambda}^i\gamma^\mu
D_\mu \lambda^i
+\frac{1}{2}\d_\mu\varphi_{\alpha}\d^\mu \varphi^\alpha+i \lambda^i[\lambda^j,(\sigma_\alpha)^{ij}\varphi^a]\right.\nonumber \\
&&\left.+i
\bar{\lambda}^i[\bar{\lambda}^j,(\sigma^*_\alpha)^{ij}\varphi^a]+\frac{1}{4}
[\varphi_\alpha,\varphi_\beta][\varphi^\alpha,\varphi^\beta]+\sum_{n_i=1}
{\cal{L}}^{KK}_{n_1...n_4}\right) \ee where by
${\cal{L}}^{KK}_{n_1...n_4}$ we collectively denote all massive
Kaluza-Klein contributions. In addition, we have combined the 3
original scalars $\phi^i$ and the 3 scalars $(A_4,A_5,A_6$)
originating from the DR of $A_M$ in
$\varphi_{\alpha}=(\phi_i,A_{3+i})$.

Now let us consider the $B_2=T^3/\mathbb{Z}_2$ compactification.
This amounts in shifting certain modes from the massless to the
massive sector of the 4D theory. With an expansion of the form
(25,27), the 4D theory turns out to be as above but with an
additional mass term \be S^{(1)}_{\rm eff}= S_{\rm eff}+\int d^4x
\frac{1}{2}{\rm Tr}\, M_{\alpha\beta}\varphi^a\varphi^\beta
\label{swe} \ee The existence of the mass term clearly breaks
susy. Indeed, there are interactions  missing from the 4D
effective theory (\ref{swe})  on $B_2=T^3/\mathbb{Z}_2$. Written
in ${\cal{N}}=1$ language, the superpotential is \be
W=\frac{1}{3}\epsilon^{ijk}\Phi_i\Phi_j\Phi_k
+M_{ij}\Phi^i\Phi^j\, , ~~~ i,j,k=1,2,3 \ee where we have define
$\Phi_i=A_{3+i}+i \phi_i$. Then clearly, the interactions from
$\lambda\d^2 W/\d\Phi^2\lambda$ \be \lambda \lambda M \Phi \ee are
missing from the effective action (\ref{swe}). Depending on the
form of the mass term in  (\ref{swe}), the ${\cal{N}}=4$
supersymmetry can either break to ${\cal{N}}=1,0$. Thus, the
$B_2=T^3/\mathbb{Z}_2$ compactification of the 7D ${\cal{N}}=2$
theory is described by an effective 4D theory with
non-supersymmetric interactions among the fields.

At this point let us compare supersymmetry breaking  described above to the one obtained through the Scherk-Schwarz mechanism.
According to the latter, employing the $R$-symmetry of the theory, one may give masses to certain fields such that supersymmetry may be broken.
In a $S^1$ compactification, one may impose the condition
\be
\Phi(x^\mu, y+2\pi L)=e^{2\pi\, i Q_\Phi}\Phi(x^\mu)
\ee
where $Q_\Phi$ is the $R$-charge of the field $\Phi$.
This leads to splitting of the $4D$ masses of the various fields according to their $R$-charge. Fermions and bosons, having different
$Q_\Phi$, obtain different contributions to their masses and supersymmetry is broken. This looks much like our boundary conditions (\ref{bd}) or (\ref{Psi}).
However, as gauge fields $A_M$ are always $R$-singlets, (vectors never carry $R$-charge, except when the $R$-symmetry is gauged), it is not possible to
aquire modified  boundary conditions. Vector fields, as well as higher-rank tensors, have $Q_\Phi=0$ and obey periodic boundary conditions under
 translations
in the extra dimension. This should be contrasted to our case, where, due to the rotation in the $x,\,y$ plane involved, vectors, as well as higher-rank
tensors, do not necessarily obbey periodic boundary conditions, as we have already seen.
As a result, in spite of their similarities, box compactification
and SSSB are different. It should also be noted that the profile of our {\text{box compactification}}
is that of spontaneous breaking with a vanishing supertrace, a
feature not shared by SSSB as the latter breaks globall supersymmetry explicitly where the mass-square
supertrace is not necessarily zero. We have also to stress that there is no way to make all components of a vector periodic due to non-homogeneity of the
box, which is manifest exactly in the different $k_3$-periodicity of the $A_M$ components.

Although in this paper the emphasis has been given to the breaking
of supersymmetry, box compactification can equally well lead to
gauge symmetry breaking. This may be discussed independently from supersymmetry and, thus,
we will consider for example an $SU(5)$ gauge theory in $7D$.
After compactifying on $B^3$,  we may expand the $7D$ gauge
fields $A_M^I,~I=1,\ldots 24$ in terms of the  $B^3$ harmonics as we did above.
 We can exploit our freedom to choose the boundary
conditions and take \be &&A_\mu^I=A_\mu^I(x)Y_{\{n\}}^{(+)} ~~~~~
\mbox{for $I$ in
$SU(3)\!\times\! SU(2)\!\times\! U(1)$}\nonumber \\
&&A_\mu^{I}=A_\mu^I(x)Y_{\{n\}}^{(-)} ~~~~~  \mbox{otherwise}
\ee
Then, clearly, the fields $A_\mu^i(x)$ have a massless mode, identified with the usual $4D$
gauge bosons, while all the rest $X,Y$ bosons are massive. However,
we also get the scalars $A_m^I$ which we should make massive by
choosing $ A_m^I=A_m^I(x)Y_{\{n\}}^{(-)}$.

Similarly for a Higgs in the fundamental $H^A, A=1,\ldots,5$ we
may take
\be
&&H^A=H^A(x)Y_{\{n\}}^{(+)}~~~~~  \mbox{for $A$ in
$SU(2)$}\nonumber \\
&&H^A=H^A(x)Y_{\{n\}}^{(-)}~~~~~  \mbox{otherwise}
\ee
The above expansions at this stage look rather \textit{ad hoc}.
 The following can serve as a hint of how they could arise.
Assume that the $\mathbb{Z}_2$ symmetry acts also in the gauge sector as
\be
\mathbb{Z}_2\subset U(1)\subset SU(5)~~:~~~ g \, {\bf 5}=-{\bf 5}\, ~~~~g \,{\bf 24}=+{\bf 24}
\ee
for the fundamental ({\bf 5}) and adjoint ({\bf 24}) of $SU(5)$. In other words, we embed
$\mathbb{Z}_2$ in the $U(1)$ subgroup of $SU(5)\supset SU(3)\!\times\!SU(2)\!\times\! U(1)$ and
we assign periodic and anti-periodic $\mathbb{Z}_2$-``parity" to the adjoint and fundamental reps, respectively.
Then, in the branching
\be
{\bf 5}=({\bf 2,3})_3+({\bf 1,3})_{-2}\, , ~~~~~{\bf 24}=({\bf 1,1})_0+({\bf 3,1})_0+({\bf 1,8})_0+({\bf 2,3})_{-5}
+({\bf 2,\bar{3}})_5
\ee
we have to choose periodic (+), or anti-periodic (-) boundary conditions according to their $U(1)\,{\rm mod2}$-charge.
Thus, for a Higgs in the fundamental, the triplet will have antiperiodic boundary conditions and, thus,
 it will have no massless mode, while the doublet will be periodic and will have a massless mode.
In contrast, for the adjoint, the $({\bf 2,3})_{-5}$ and $({\bf 2,\bar{3}})_5$ will have
no massless mode, as they have odd
$U(1)\,{\rm mod2}$-charge and the $\mathbb{Z}_2$-``parity" of the adjoint is $+1$.

The recent activity on theories and models characterized with large extra dimensions provides a framework that can acommodate a connection between
the phenomenologically required small supersymmetry breaking and compactification. In the present short article we analyzed the basic features of a
novel compactification scheme on a flat three dimensional torus, where opposite sides are identified after two of them have undergone a rotation by
$\pi$. Although the scheme superficially resembles orbifold compactification it is not an orbifold compactification, since it
does not involve any fixed points. Starting with a supersymmetric theory, the chosen boundary conditions for component fields can be such that
lead to a compactified theory with reduced or completely broken supersymmetry. Examples of boundary conditions that, for a $7D$ theory, lead to
$N=4\rightarrow N=2,\,N=1,\,N=0$ breakings were worked out. It remains to be seen in future work whether this framework can be used for the
construction of realistic models. The spectrum profile of the supersymmetry breaking scheme discussed is analogous to the one associated with
spontaneous supersymmetry breaking, characterized by a vanishing supertrace. We should also stress once more the difference of the present scheme to
the Scherk-Schwarz supersymmetry breaking scheme in which component fields aquire non-trivial boundary
 conditions through their different $R$-symmetry charges. In this scheme vector fields cannot be affected. In contrast, here the compactification
 scheme allowes for non-trivial gauge field boundary conditions. Although, we did not elaborate much on gauge symmetry breaking, it is clear that
 box compactification can naturally serve as a way to break gauge symmetries as well in ways analogous to the ones employed in orbifold
 theories \cite{VW}. An
 intriguing question not touched by the present first short presentation of box compactification is that of the arbitrarines of the chosen boundary
 conditions. The answer is linked to the quantum dynamics that will ultimatelly discriminate between the various available compactification
  solutions.

\end{document}